# PDBMine: A Reformulation of the Protein Data Bank to Facilitate Structural Data Mining

Casey A. Cole
Department of Computer Science
University of South Carolina
Columbia, SC, USA
coleca@cse.sc.edu

Christopher Ott
Department of Computer Science
University of South Carolina
Columbia, SC, USA
caott@email.sc.edu

Diego Valdes
Department of Computer Science
University of South Carolina
Columbia, SC, USA
dvaldes@email.sc.edu

Homayoun Valafar
Department of Computer Science
University of South Carolina
Columbia, SC, USA
hvalafar@gmail.com

*Abstract—* Large scale initiatives such as the Human Genome Project, Structural Genomics, and individual research teams have provided large deposits of genomic and proteomic data. The transfer of data to knowledge has become one of the existing challenges, which is a consequence of capturing data in databases that are optimally designed for archiving and not mining. In this research, we have targeted the Protein Databank (PDB) and demonstrated a transformation of its content, named PDBMine, that reduces storage space by an order of magnitude, and allows for powerful mining in relation to the topic of protein structure determination. We have demonstrated the utility of PDBMine in exploring the prevalence of dimeric and trimeric amino acid sequences and provided a mechanism of predicting protein structure.

## I. INTRODUCTION

Completion of the Human Genome Project in 1990[1] marked the beginning of the era of Big Data. Since then, various funding agencies initiated numerous large scale studies such as Structural Genomics Initiative[2], Protein Structure Initiative[3], and Genomic Data Commons[4] that have generated unimaginable volumes of data. Currently GO[5] houses over 7 million annotated genes and PDB[6] houses over 144,729 protein structures alone. Many other existing repositories of internationally collected data can be listed. While the advancement of technologies and scientific methods have contributed to the large growth in the data volume, velocity, and variety, the collected data has not had the anticipated impact in expansion of our knowledgebase. The limited impact of these databases is due to the fact that these repositories are optimized for data deposition but not for data mining. Recognizing this limitation, various funding agencies (including NSF and NIH) have declared new initiatives with the objective of transforming data to knowledge. Such transformation will require re-representation of data in such a manner that facilitates mining and knowledge discovery. Here we present the first instance of transformation of the Protein Databank (PDB) that allows for discovery of knowledge.

Proteins are a class of macromolecules that perform various functions in cells including structural support, enzymatic activities, cell signaling, and more. Proper regulation of proteins is crucial for life in all living organisms. Proteins are made up of smaller subunits called amino acids that are structurally defined by a series of angles called dihedrals. Large scale mining of this data can potentially lead to prediction of structure, function, binding sites, and better understanding of evolutionary relationships between organisms[7, 8].

The primary database that houses structural information of proteins, PDB, contains all the necessary information needed to probe for structural insights but is not configured in such a way as to make the task straightforward. For example, there is currently no capability to easily extract dihedrals from these coordinates for a given sequence. To extract these angles for use in structure prediction algorithms, one would need to first perform a sequence search across all proteins, download all the hits, and finally write a script, or series of scripts, to extract the coordinates from the PDB files to perform the calculation of dihedrals. Each step of this process is difficult and/or time-consuming. Our new database, PDBMine, will alleviate these difficulties and make it easy to extract information such as dihedral angles quickly and accurately from raw PDB coordinates for use in a variety of applications. In addition to various observations resulted from PDBMine, we present our its preliminary application in prediction of protein structure.

## II. BACKGROUND AND METHOD

### A. Existing Databases and Their Limitations

Protein DataBank or PDB (https://www.rcsb.org/) is historically the oldest international repository of macromolecular structures dating back to as early as 1971. Currently, PDB houses the three-dimensional coordinates of 144,729 protein structures and provides an array of search mechanisms. However, the search mechanisms provided by PDB are aimed at navigating and retrieving the contents of the database. Therefore, there is little to no capability of querying at a more fine-grain level that will allow for knowledge discovery. Over the past decade, there have been several attempts[9, 10] that aimed at creating a database for fine-grained mining of protein databases. In 2006, the DASSD[9] was created as a database that housed short protein fragments (sizes 1, 3, and 5 amino acids). Through now an inactive website, users were able to enter a sequence of amino acids to receive structural information regarding the middle

This work was funding by NIH grant number P20 RR-016461

residue of the query. In addition, DASSD would also provide a prediction related to the secondary structure formation of a given fragment. The limitation of the input size (1, 3 or 5 residues) made predictions of larger proteins implausible. Furthermore, as we demonstrate in the results, fragments of size 3 or 4 residues do not provide sufficiently converged results to meaningfully define a protein structure. Therefore, it is critical that a database to be capable of querying all fragment lengths including k-mers with k>=5. Protein Geometry Database (PGD)[10] is another attempt at creation of minable database of protein structures. This database extended the maximum fragment size to 10 amino acids and added additional search criteria such as R-factor and x-ray resolution. However, the current version of the database contains information for only 16,000 protein structures determined from only x-ray. While PGD may provide mining of "high-quality" structures, the limited number of protein structures and methods of characterization could lead to erroneous or biased results, especially when dealing with proteins that are inherently difficult for x-ray diffraction (such as membrane proteins and proteins that undergo dynamics). The database presented in this paper aims to overcome the limitations of these implementations and therefore create a more complete and encompassing platform for analyzing local and global protein geometries.

**B. Transformation of PDB into PDBMine**

Data and their corresponding databases can be formulated and transformed to reduce their space requirements, to increase data retrieval speed, to serve as more secure repository of data, to name a few example final objectives. In this work we transformed the existing PDB data with the primary goal of providing a more useful structural mining database. To that end, we extracted the following information from each protein structure: the mechanism of structure determination, the primary sequence of every protein, backbone torsion angles, hydrogen bonding, surface accessibility, as well as the three-dimensional coordinates of each atom. The derived information was then captured in PDBMine (simplified schema shown in Figure 1). The program DSSP[11] was used to convert each of the downloaded *pdb* files to their corresponding *dssp* files, which contained the surface accessibility, hydrogen bonding information, and ɸ/Ψ angles for each residue. The contents of the DSSP file was then stored in the PDBMine database. MySQL was chosen as the platform for the PDBMine and a series of Python scripts were developed to parse and store the data into the database.

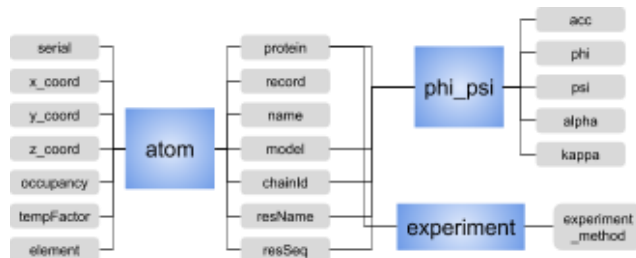

*Figure 1. Database schema for PDBMine.*

PDBMine consists of three tables: ATOM, PHI_PSI and EXPERIMENT. The ATOM table contains the protein name, residue number, residue type, atom type as well as the atomic coordinates (X, Y, Z) for every atom in every protein. Storing the protein in this raw form allows for extracting spatial information such as "finding all the carbon atoms that are within 2Å of nitrogen" or "finding all the alanines that are within 5Å of a proline." Such queries could be used to identify proteins with certain structural motifs. The PHI_PSI table contains the dihedrals for all residues of each protein as well as other information collected from the DSSP program such as surface area accessibility. Finally, the EXPERIMENT table summarizes the metadata associated with the protein such as the experimental method of structure determination.

**C. Application of PDBMine in Structure Prediction**

One application of the PDBMine database is prediction of backbone dihedral angles of a protein. Three dihedral angles phi (ɸ), psi (Ψ), and omega (Ω) define the structural variability of the protein backbone at each amino acid. The collective effect of these torsion angles defines the overall structure of a protein. Due to the biophysical properties of the peptide bond, the Ω torsion angle is generally fixed at 180° (or occasionally 0°). Therefore, the effective degrees of freedom at any given amino acid are ɸ and Ψ. Current methods of protein structure prediction and calculation[12, 13] rely on, at least to some extent, accurate prediction of dihedral angles for a given set of amino acids or a k-mer. These "local" dihedral predictions are used as scaffolding for the prediction of the full global structure. It, therefore, becomes important for the local k-mer geometries to be accurately predicted. If they are inaccurate then stitching the k-mers together to create the global structure will produce erroneous results.

Here we present an example application of PDBMine to facilitate more sophisticated and complete data analytics of the protein backbone dihedrals. In this application, we have created a frontend to the PDBMine with the specific task of collecting, analyzing, and reporting of the data. More specifically, this frontend accepts a protein primary sequence, and a search window size of k (where n >=k). The protein sequence is then automatically dissected into k-mers using a rolling window. Each k-mer is then queried in the database. The results for each of the queries are collected into a series of CSV files that contain the PDB accession number of the database hit, the chainID, model number, amino acid name, and the corresponding ϕ/Ψ angles. Furthermore, the backbone torsion angles for each residue is consolidated by combining the results for every one of the k places that the amino acid could appear in a k-mer rolling window. Finally, using the aggregated dihedrals for each residue and Kernel Density Estimation[14, 15], the most likely dihedral is predicted. All final and intermediate results are compiled and sent to the user via email.

### D. Evaluation Techniques

In 1963, G.N. Ramachandran noted the seminal observation that the ϕ/Ψ values in proteins adhere to a more restricted range of angles[16]. This restricted space of protein backbone dihedrals is denoted as Ramachandran plot (or R-Space). The first step in the evaluation of PDBMine was to query and recreate the previously reported R-Space[17-19] for each amino acid. The resulting distribution plots are then compared to the previously published and well accepted R-Spaces for each of the amino acids. This step will serve as a validation step through agreement with the previously reported work. As an extension, more complex and novel (previously unknown) R-Spaces were also created from extended k-mers (2-mers and 3-mers). The novel R-spaces serve as examples of new information that can be produced from mining the PDBMine. Finally, we have demonstrated the potential of PDBMine in application to the challenging task of protein structure prediction. In this context, using the results from the database, the protein structure of ubiquitin was predicted purely based on statistical sampling of the backbone dihedrals. The predicted structure was compared to the x-ray structure currently published in PDB.

## III. RESULTS AND DISCUSSION

### A. Database Creation

The total time consumed for downloading, parsing and uploading all proteins within the PDB was ~2016 hours, or 84 days. The final total space requirement for the database was 310 GB. This is an improvement over the space that is currently required (over 1 TB) to store the protein structures in pdb format in the PDB.

Of the 144,729 number of protein structures that were parsed, 3,764 of them required additional treatments due to file abnormalities such as the presence of DNA and RNA molecules, missing atoms, misnamed atoms, and others. These anomalies were addressed by designing and deploying specific scripts, after which, the final product was parsed and uploaded to PDBMine.

### B. Results of the Data Mining and Analysis

*Evaluation of Data* - As a prerequisite step, some basic analyses were performed to validate the content of PDBMine based on previously known information. The first of which was to calculate the abundance of each single amino acid and compare it to the statistics published from UniProt[20] (a database of all known protein sequences). Figure 2 shows a comparison of amino acid abundance from the two sources in a grouped bar chart. The red and blue bars represent the calculated percentage occurrence of each amino acid in PDBMine and UniProt respectively. The two figures demonstrate very close agreement between two sources, indicating validity of the PDBMine's data. In this figure, the largest observed difference is for the amino acid tyrosine (Y).

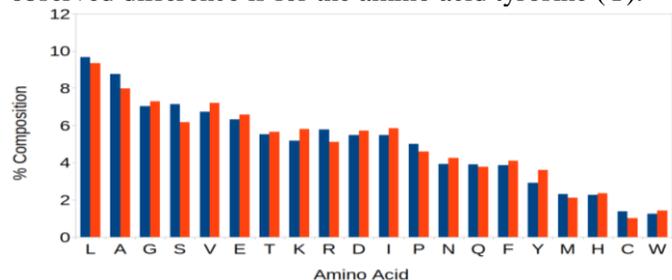

*Figure 2. The abundance of each amino acid found in Uniprot (blue) and PDBMine (red).*

The abundance of amino acid 2-mers and 3-mers was also mined from PDBMine. Figure 3 shows the percentage appearance of all 2-mers (400 combinations) in all known protein structures. Although it is difficult to gleam the exact count number for a given 2-mer from the figure, it is included here to show the general shape of the distribution. In particular, to illustrate that not all dimers are uniformly present. For instance, four dimers (LL, AL, AA, LA) occur ~700,000 times while three dimers (CW, CC, WC) only occur ~15,000 times (nearly 50 times less).

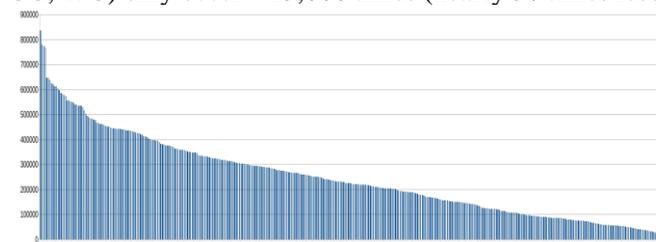

*Figure 3. The general shape of the distribution of 2-mers in the database.*

The percentage occurrence of 3-mers follows the same general shape as the 2-mer distribution shown in Figure 3. The general shape of the distribution of 2-mers in the database.Figure 3. There are six 3-mers (ALL, EAL, ALA, AAL, LAA, AAA) that occur over 70,000 times within the database whereas there are eight (CHW, MWC, CMW, CCW, HWC, CIW, WWC, WCM) that occur less than 200 times. At the time of submission, there were no publications that detailed reasons as to the significance of some k-mers occurring more often than others.

In addition to comparing the distributions of amino acids, the R-Space was extracted for each of the 20 amino acids. These were visually compared to the known, accepted R-Spaces for single amino acids. For brevity, only the case of GLY and PRO are presented as they have R-Spaces that differ significantly from the other 18 amino acids. Figure 4a shows the comparison of the PDBMine generated (left) and accepted[21] (right) R-Spaces for GLY. Part b of Figure 4 depicts the same for PRO.

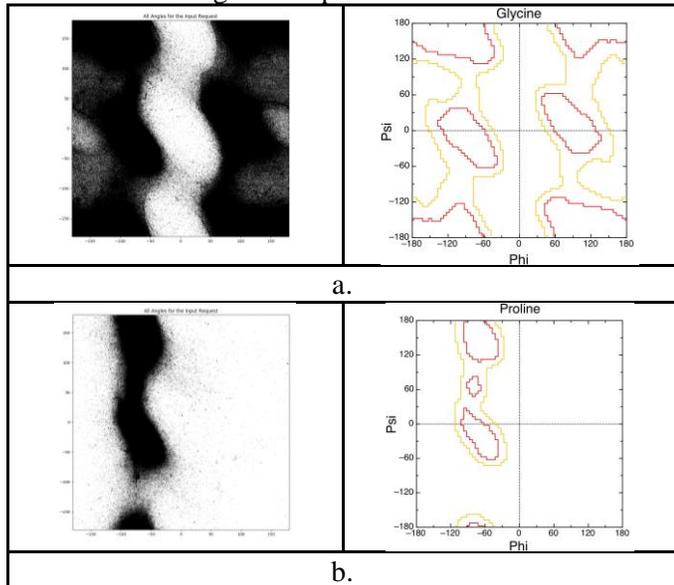

Figure 4. R-Space (PDBMine on left and accepted on right) for a) GLY and b) PRO.

*Prediction of φ/Ψ Angles* - The utility of PDBMine can well exceed beyond the scope of a single amino acid. Figure 5a-b offer some useful insights for the common amino acid pair of glycine-proline. This motif occurs often in protein structures especially in the loop and turn regions. Figure 5a shows the R-Space for the glycine in the context of the glycine-proline combination. Proline is a relatively rigid amino acid whereas glycine is very flexible due to their respective sidechain configurations. In comparison to the typical glycine space (Figure 4a), Figure 5a shows a much more restrictive area of permissible torsion angles. Figure 5b depicts the R-Space for the proline of all glycine-proline amino acid pairs. In this case, the addition of the glycine does not change the R-space significantly for proline. Figure 5c-d show the R-Space for proline-proline pairs. In this pairing, the φ/Ψ angles for the first proline (panel c) are significantly restricted compared to the traditional R-Space shown in Figure 4b. The second proline in the pair, shown in panel d, however, shows much better agreement with the traditional proline R-space. The proline-proline motif occurs in proteins fairly often with a current count size of 204,994 and, therefore, an increased understanding of its local structure will be of great benefit to computational methods.

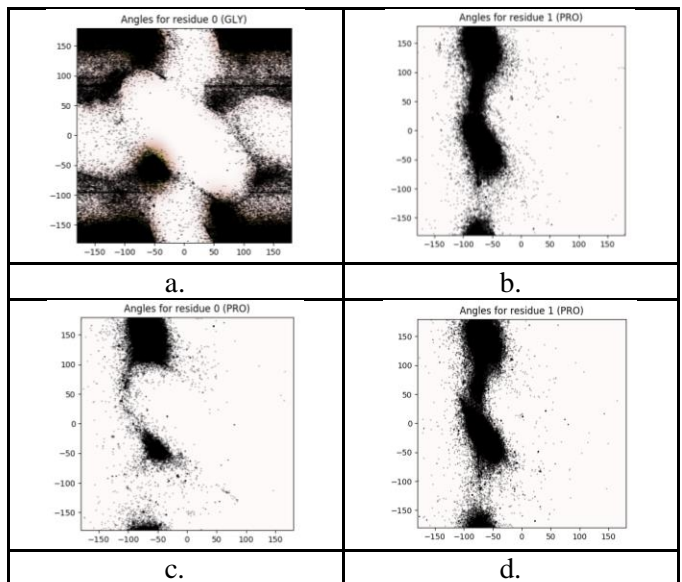

Figure 5. R-Spaces for a) GLY of GLY-PRO, b) PRO of GLY-PRO, c) first PRO of PRO-PRO and d) second PRO of PRO-PRO.

In addition to calculating the R-space for amino acids and 2-mers, this method can be extended to k-mer φ/Ψ prediction. One example is shown in Figure 6 depicting the R-Space for the 3-mer GLY-PRO-PRO. As it can be seen the space of allowed dihedral angles are significantly more limited compared to Ram-Space of a single amino acid (compare Figure 6a-c to Figure 4a,b).

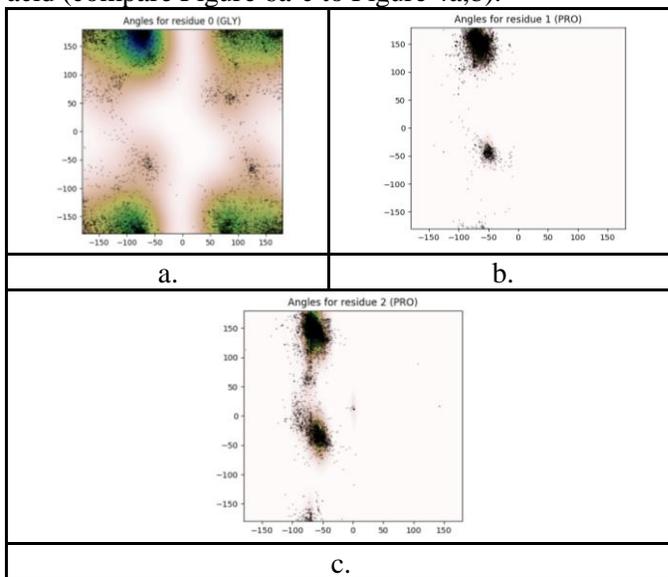

Figure 6. R-Space for the triplet GLY-PRO-PRO a) for GLY, b) for first PRO, c) for second PRO.

*Protein Structure Prediction* - The protein ubiquitin (76 residues) has been the subject of numerous studies by both experimental and computational methods[18, 22] of structure calculation. This makes it an ideal candidate for a proof-of-concept case. The dihedral angles of ubiquitin were calculated using a KDE-based prediction of k-mer dihedrals with k values of 3, 6, and 7. Examples of deviation in R-spaces for a given amino acid is shown in Figure 7. Notice that as the k increases (from left to right), the dihedral space becomes increasingly confined which leads to, as shown later, better structure prediction.

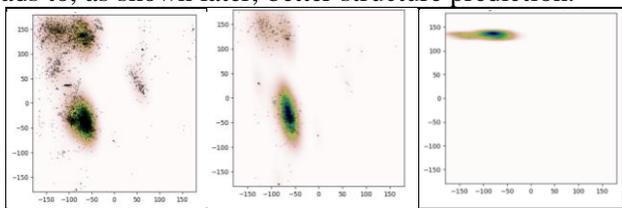

*Figure 7. Examples of the differences for residue 14 of ubiquitin at lengths of k= 3, 6 and 7*

Structures were generated based on the results for the three different experiments (3-mer, 6-mer, 7-mer) using the program "pdbgen" included in the REDCRAFT[17, 23-27] software package. The structure using k=3 (shown in Figure 8 in red) exhibited a backbone root mean squared deviation (bb-RMSD) of over 22Å to the crystal structure PDB-ID:1UBQ[28]. This indicates a low level of overall structural similarity. The resulting structures for 6 and 7 (shown in green and purple respectively in Figure 8) were similar with both exhibiting a bb-RMSD of around 3.5Å to the known crystal structure. This bb-RMSD indicates a reasonably high level of similarity between the two structures.

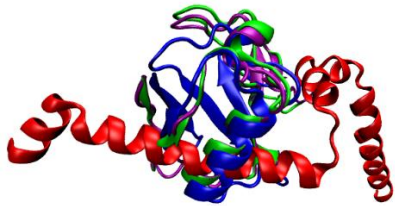

*Figure 8. Resulting structures for k=3,6,7 (red, green and purple) aligned to the x-ray structure (blue).*

Further analysis using multiple structure alignment software msTALI[7, 8] showed that the conserved core between these three structures (k=6,7 and the x-ray) contained 57 residues. The remaining 19 residues contributed to the divergence in structure (bb-RMSD). It is also worth noting, that a structural alignment including the result from k=3 yielded a core conserved region of 25 residues. While this indicates more regions of divergence, it also illustrates the amount of valuable information that is still present even at the k=3 level.

**C. Web Resources**

A preliminary interface to the database has been created that will allow for fast, easy extraction of dihedral angles given a k-mer (**ifestos.cse.sc.edu/frontend**). The website was created using HTML/CSS, while the backend computation and data storage is accomplished using a combination of Python scripts and a mySQL database.

*Navigation* - The first page on the website the user inputs a window size, amino acid sequence, and an email address. After making a submission the user is provided with a summary of their input or an error message.

*Usage* - To submit a query, the user needs to provide a window size, an amino acid sequence, and a contact email address. The sequence can be submitted either as a single amino acid string (E N I E... etc) or in a triplet format (GLU ASN ILE GLU... etc). For the triplet format, the user needs to check the "Use AAA Formatting" option. Once submitted, the server will schedule the query and return the results once complete. These results contain a list of the predicted maximum likelihood angles for each residue in the request, plots of the KDE for each residue, and folder of the CSV files returned by the database. These CSV files contain a list of all proteins in the database that contained that k-mer along with the dihedral measures for each in which the user can perform their own additional analyses on the dihedral information.

*Capabilities* - Our local, fragment-based approach allows the user to obtain predicted structural information for sequences that have low global similarity with existing entries in the PDB. Changing the window size allows the user to control the amount of data returned. Larger sizes are more restrictive but will produce more well-defined results. Smaller sizes can be employed for sequences with unusually low representation in the PDB.

**D. Future Work**

Future work will include improvements in two major areas: angle prediction and the web interface. Machine learning techniques such as traditional and deep neural networks can be used in place of the KDE method to improve the prediction of dihedral angles. Additions to the website will include advanced filtering including the ability to select only proteins characterized by certain experimental methods as well as the ability to select certain PDB ids to be excluded. In addition to these advances in capability, there will be additional graphical changes including onsite interactive visualization of R-Spaces for a given k-mer as well as automatic generation of protein structure ensembles from predicted angles.

## IV. CONCLUSION

In this work, PDBMine, a database of dihedral angles mined from known protein structures, was presented. To demonstrate the validity of the data, known R-Spaces were generated and compared to their respective counterparts. In addition, preliminary results were shown for protein structure calculation using solely KDE-based prediction of dihedral angles. The web interface of this database allows for easy and efficient retrieval and analysis of dihedral angles for k-length amino acid sequences. The output of this website can be easily incorporated into existing protein structure calculation tools for increased accuracy of models. In future work, more sophisticated mechanisms of prediction will be utilized, and improvements will be made to the web interface to allow for more flexible querying.